\documentclass{PoS}

\newcommand{\SU}{\mathrm{SU}}

\newcommand{\Gtwo}{\mathrm{G}_2}

\newcommand{\dd}{{\rm{d}}}

\newcommand{\THagedorn}{T_{\mbox{\tiny{H}}}}
\newcommand{\TNG}{T_{\mbox{\tiny{NG}}}}

\newcommand{\Up}{U_{\mbox{\tiny{p}}}}

\newcommand{\eq}{\begin{equation}}
\newcommand{\en}{\end{equation}}
\newcommand{\eqar}{\begin{eqnarray}}
\newcommand{\enar}{\end{eqnarray}}

\title{Universal aspects in the equation of state for Yang-Mills theories}

\ShortTitle{Universal aspects in the equation of state for Yang-Mills theories}

\author{\speaker{Alessandro~Nada}\\
	Dipartimento di Fisica, Universit\`a di Torino and INFN, Sezione di Torino\\
	Via Pietro Giuria 1, I-10125 Torino, Italy\\
	E-mail: \email{anada@to.infn.it}}

\abstract{We present high-precision lattice calculations of the thermodynamics of Yang-Mills theories with different gauge groups. 
In the confining phase, we show that the equation of state is described remarkably well by a gas of massive, non-interacting glueballs, provided that an effective bosonic closed-string model is used to derive an exponentially growing Hagedorn spectrum for the heavy states. 
In particular, this model describes very accurately the results for the SU(3) theory reported by Bors\'anyi et al. in JHEP 07 (2012) 056, as well as a novel set of lattice data for the SU(2) theory. 
In addition, we also also show that the equation of state in the deconfined phase exhibits a near perfect proportionality to the number of gluon degrees of freedom, including for the Yang-Mills theory based on the exceptional, center-less gauge group $\mathrm{G}_2$.}

\FullConference{The European Physical Society Conference on High Energy Physics\\
                 22-29 July 2015\\
                 Vienna, Austria}

\begin{document}

\section{Introduction}

Pure glue non-Abelian gauge theories retain many features of real-world QCD and reveal precious insight that is applicable, at least on a qualitative or semi-quantitative level, to the full theory. In addition, it has been recently pointed out that they may even be of direct phenomenological relevance for the description of the early stages of the matter produced in high-energy nuclear collisions~\cite{Stoecker:2015zea}.
In this contribution we discuss recent progress in this line of research aimed at exploring the finite temperature behaviour of different pure gauge theories.
In the first part we present a recent work~\cite{Caselle:2015tza} on the equation of state (e.o.s.) in the confining region of the SU(2) Yang-Mills gauge theory, which compares a novel set of lattice results with the prediction of a non-interacting glueball gas.
A very good agreement between the two is found, provided that a closed bosonic string model is used to describe the high-lying glueball states contribution to the glueball gas, as first proposed in a seminal lattice study by Meyer~\cite{Meyer:2009tq}.
This string-inspired approach is successful also in describing with no free parameters previous lattice results for SU(3) from ref.~\cite{Borsanyi:2012ve}: as explained in the final section, the comparison with the $N=2$ theory provides a critical cross-check for the model.
In the second part results from a recent work~\cite{Bruno:2014rxa} on the e.o.s. of the exceptional $\Gtwo$ Yang-Mills gauge theory in the deconfined phase are reviewed.
The interest for pure-glue $\Gtwo$ lies both in the similarities with full QCD, for example dynamical string-breaking at large distances (but due to gluons), and in the differences with SU($N$) gauge theories, namely the fact that the center of $\Gtwo$ is trivial.
Results from lattice simulations in the $T \lesssim 3T_c$ region are presented; in this temperature range it is possible to point out some analogies with SU($N$) gauge theories, supporting the idea of a universal thermal behaviour between different confining gauge theories.

\section{Thermodynamics on the lattice}

The pure-glue Yang-Mills gauge theories considered in this work are regularized on a four-dimensional hypercubic lattice $\Lambda$ of spacing $a$ and hypervolume $a^4(N_s^3 \times N_t)$, with periodic boundary conditions on all directions.
The lattice spacing $a$ is a function of the Wilson parameter $\beta$, defined as $\beta=2N/g^2$ for SU($N$) theories and as $\beta=7/g^2$ for the $\Gtwo$ theory, with $g^2$ being the bare squared lattice coupling.
The physical temperature of the system is given by the inverse of the shortest (temporal) size, i.e. $T=1/(a(\beta)N_t)$: in these works variations of the temperature are performed by changing the lattice spacing $a(\beta)$ while keeping $N_t$ fixed.

A thermodynamic quantity of major interest in finite-temperature field theory is the pressure $p$, which in the thermodynamic limit equals the opposite of the free-energy density$f$:
\begin{equation}
\label{pressuretdyn}
p = -\lim_{V \to \infty} f = \lim_{V \to \infty} \frac{T}{V} \ln{Z}.
\end{equation}
We introduce also the trace of the energy-momentum tensor $\Delta$, which is also called trace anomaly, as it is related to the breaking of conformal invariance. 
It can be calculated as a derivative of the pressure
\begin{equation}
\frac{\Delta}{T^4} = T \frac{\partial}{\partial T} \left( \frac{p}{T^4} \right).
\end{equation}
Quantities such as the energy density $\epsilon$ and the entropy density $s$ can be easily evaluated:
\begin{equation}
\epsilon = \frac{T^2}{V} \left. {\frac{\partial \ln Z}{\partial T}}\right|_V = \Delta + 3p , \qquad s = \frac{\epsilon}{T} + \frac{\ln Z}{V} = \frac{\Delta + 4p}{T}.
\end{equation}
Following the ``integral method''~\cite{Engels:1990vr} it is possible to express the aforementioned thermodynamic quantities in terms of plaquette expectation values.
For example, the pressure at a given $\beta$ is
\begin{equation}
\label{pressure1}
p = \frac{6}{a^4} \int_{\beta_0}^{\beta} \dd \beta' \left( \langle \Up \rangle_T - \langle \Up \rangle_0 \right)
\end{equation}
where $\beta_0$ corresponds to a temperature close to zero, while $\langle \Up \rangle_T$ is the average plaquette at a generic temperature $T$.
The determination of the trace of the energy-momentum tensor is even more straightforward, as it can be expressed in units of $T$ as
\begin{equation}
\label{trace1}
\frac{\Delta}{T^4} = 6 N_t^4 \frac{\partial \beta}{\partial \ln a} \left( \langle \Up \rangle_0 -  \langle \Up \rangle_T \right),
\end{equation}
where the factor $\partial \beta/\partial \ln a$ is to be evaluated from the scale setting.
All expectation values have been estimated numerically via Monte Carlo numerical integration on a large set of configurations generated by a mix of ``heat-bath'' and ``overrelaxation'' Monte Carlo algorithms.

\section{SU($N$) theories in the confining phase and comparison with a string model}

\begin{figure}
\begin{center}
\includegraphics*[width=0.7\textwidth]{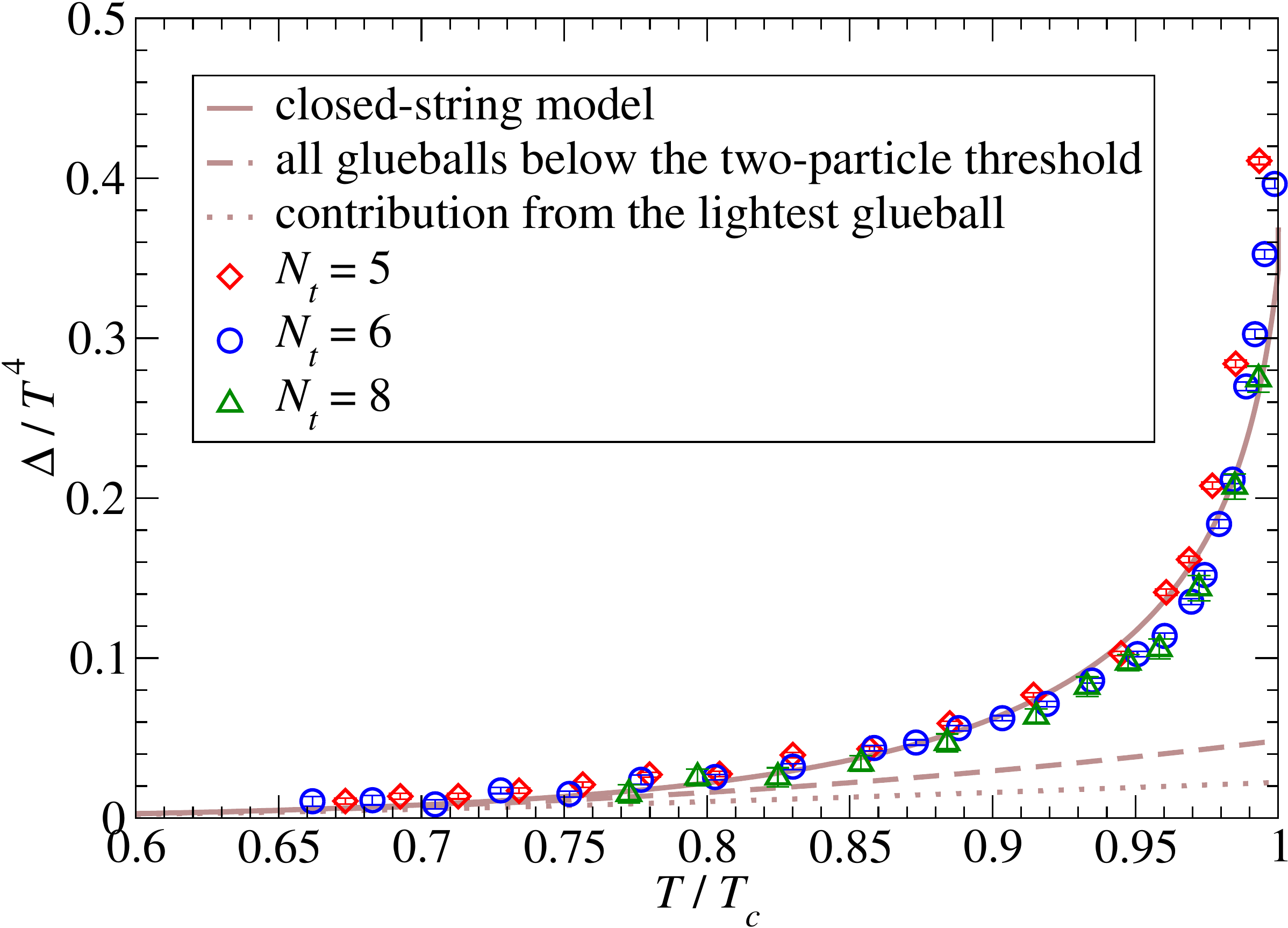}
\caption{\label{fig:su2_trace} Comparison between our lattice results for the trace anomaly in $\SU(2)$ Yang-Mills theory and the behaviour a gas of free, massive glueballs with different contributions from the spectrum.
The data are plotted using the parameter $T_c/\sqrt{\sigma}=0.7091(36)$ from ref.~\cite{Lucini:2003zr}, while the glueball gas predictions are calculated using masses from ref.~\cite{Teper:1998kw}.}
\end{center}
\end{figure}

In the first part of this article we review a novel calculation~\cite{Caselle:2015tza} of the equation of state for the SU(2) pure-glue gauge theory in the confining phase.
The results for the trace anomaly computed numerically on lattices with different temporal extents ($N_t=5,6,8$) are reported in figure~\ref{fig:su2_trace}.

Our first assumption is that in the $T<T_c$ region glueballs are weakly interacting with each other and, thus, the thermodynamics of the system can be effectively described by a gas of non-interacting glueballs.
In particular, for a free, relativistic Bose gas the trace anomaly is given by
\begin{equation}
\Delta = \frac{m^3T}{2 \pi^2} \sum_{n=1}^\infty \frac{K_1 \left( nm/T \right)}{n}.
\label{B2}
\end{equation}
The contribution to the dimensionless trace $\Delta/T^4$ of a glueball gas which includes only the lightest glueball (the $0^{++}$ state) is reported in figure~\ref{fig:su2_trace}, along with the contribution that accounts for all glueball states with mass smaller than $2m_{0^{++}}$.

\begin{figure}
\begin{center}
\includegraphics*[width=0.7\textwidth]{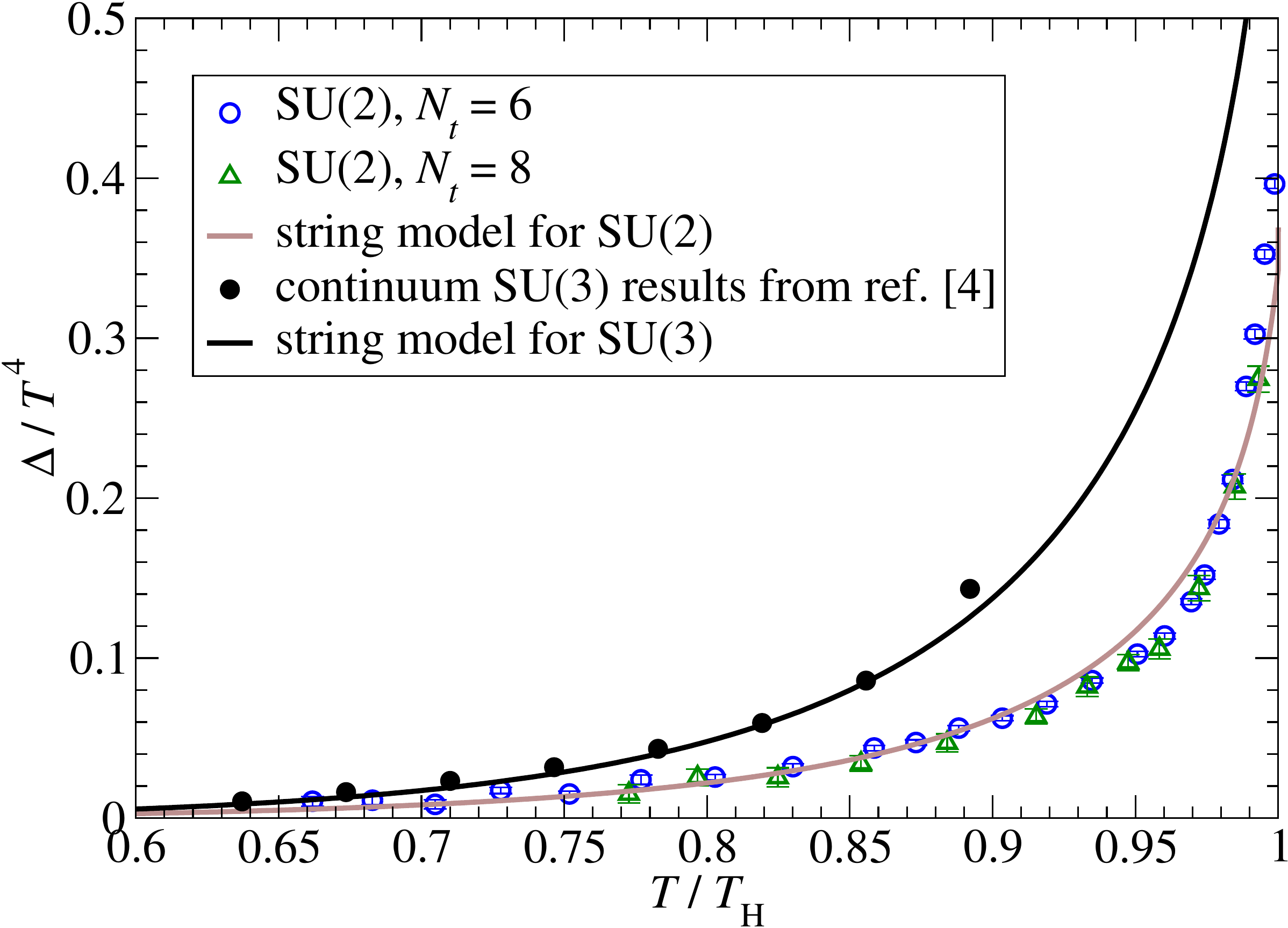}
\caption{\label{fig:su2su3} Comparison between the predictions of a massive-glueball gas, including the contribution from states modelled by a closed Nambu-Got\={o} string model, 
and continuum-extrapolated data obtained in ref.~\cite{Borsanyi:2012ve} for $\SU(3)$ Yang-Mills theory and our data for the $\SU(2)$ theory, as a function of $T/\THagedorn$.}
\end{center}
\end{figure}

It is clear that these glueball-gas predictions fail to describe correctly the lattice results for temperatures in the proximity of the deconfining transition as a large mismatch appears.
This problem can be addressed if an exponentially increasing spectrum is introduced in order to account for the heavier glueball states: one possibility to do so is by modelling glueballs as closed flux tubes which can be described as closed bosonic strings.
Indeed, a direct consequence of the bosonic string model is the existence of a Hagedorn-like exponential spectrum described by the spectral density
\begin{equation}
\label{spectraldensity}
 \hat{\rho} (m) = \frac{1}{m} \left( \frac{2\pi \THagedorn}{3m}\right)^{3} \exp \left( m/\THagedorn \right).
\end{equation}
The only parameter is the Hagedorn temperature $\THagedorn$, which for $N=2$ colors is fixed to be $\THagedorn = T_c$, as the deconfinement transition is of second order.
Using eq.~(\ref{spectraldensity}) to account for all the glueball states with masses larger than the threshold $2m_{0^{++}}$ we obtain the solid line of figure~\ref{fig:su2_trace}, which shows an excellent agreement up to temperatures very close to the transition.

It is possible to repeat the same analysis for the SU(3) pure-glue theory, which presents two crucial differences with respect to SU(2).
Firstly, the deconfinement transition is of first order, and the Hagedorn temperature must be determined somehow. If the effective string action governing the model is identified with the Nambu-Got{\={o}} action, then the value of $\THagedorn$ is predicted to be:
\begin{equation}
\label{Hagedorntemp}
 \THagedorn \equiv \TNG =\sqrt{\frac{3 \sigma}{2\pi}} \simeq 0.691 \sqrt{\sigma}.
\end{equation}
Secondly, the glueball spectrum for $N=3$ colors also includes states with charge conjugation number $C=-1$, while the SU(2) theory admits only states with $C=+1$ due to the pseudo-real nature of its representations.
This means that for SU(3) a factor two must be introduced in the spectral density (\ref{spectraldensity}): the contribution of the whole glueball spectrum for SU(3) is reported in figure~\ref{fig:su2su3} along with the SU(2) one.
Also in this case, excellent agreement is found between the parameter-free effective string model prediction and lattice data. Moreover, the crucial role of the $C=-1$ glueball states can be seen in contrast with the SU(2) glueball gas curve.
This difference between $N=2$ and $N>2$ Yang-Mills theories was already noted in a previous study 2+1 dimensions~\cite{Caselle:2011fy}, of which ref.~\cite{Caselle:2015tza} represents a direct generalization. Similar attempts to describe the trace anomaly of SU(3) theory with a relativistic model for the glueball spectrum were made in refs.~\cite{Buisseret:2011fq, Megias:2014bfa}.

\section{Results for $\Gtwo$ gauge theory for $T>T_c$}

\begin{figure}
\includegraphics*[width=0.47\textwidth]{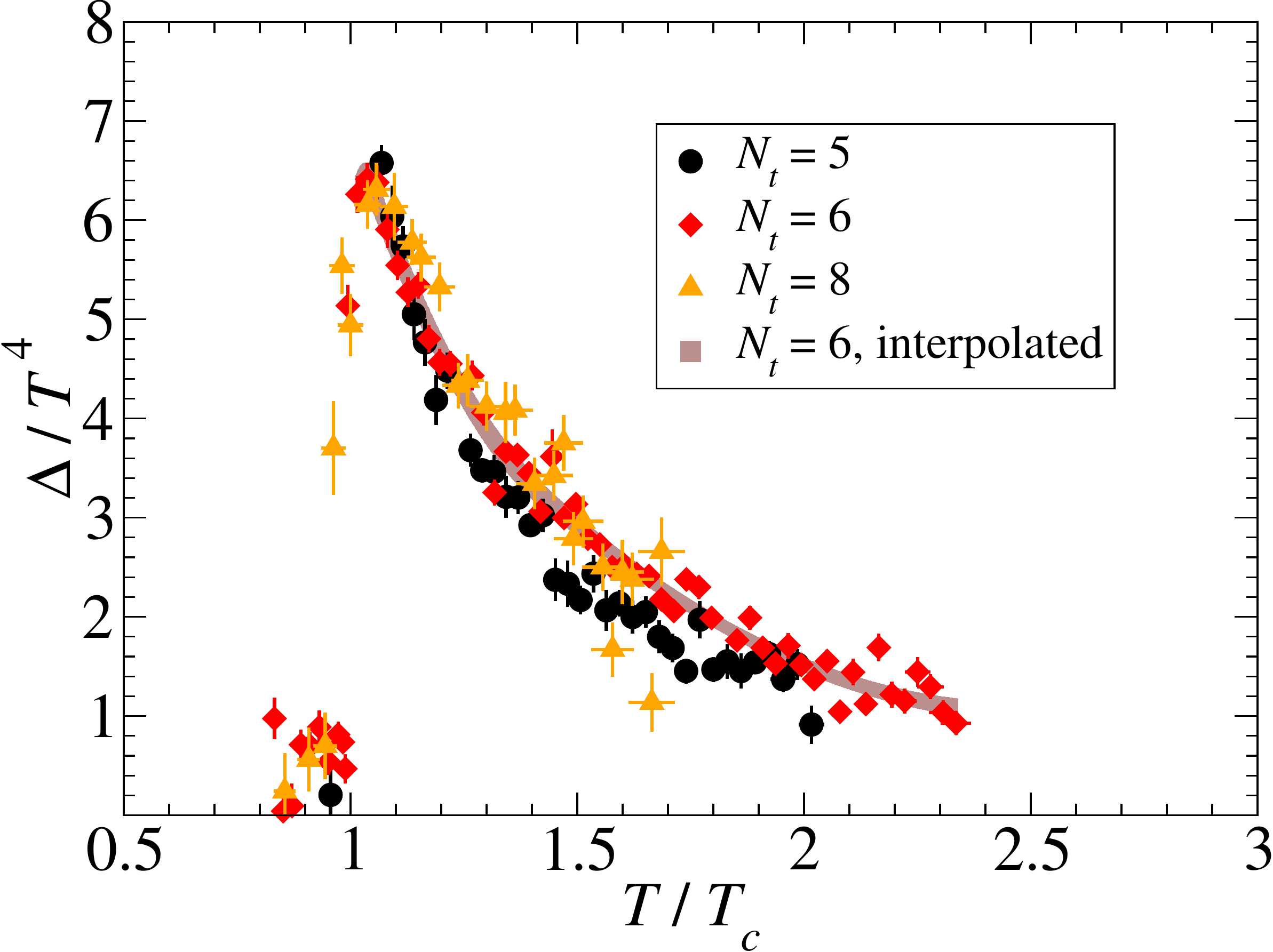} \hfill \includegraphics*[width=0.51\textwidth]{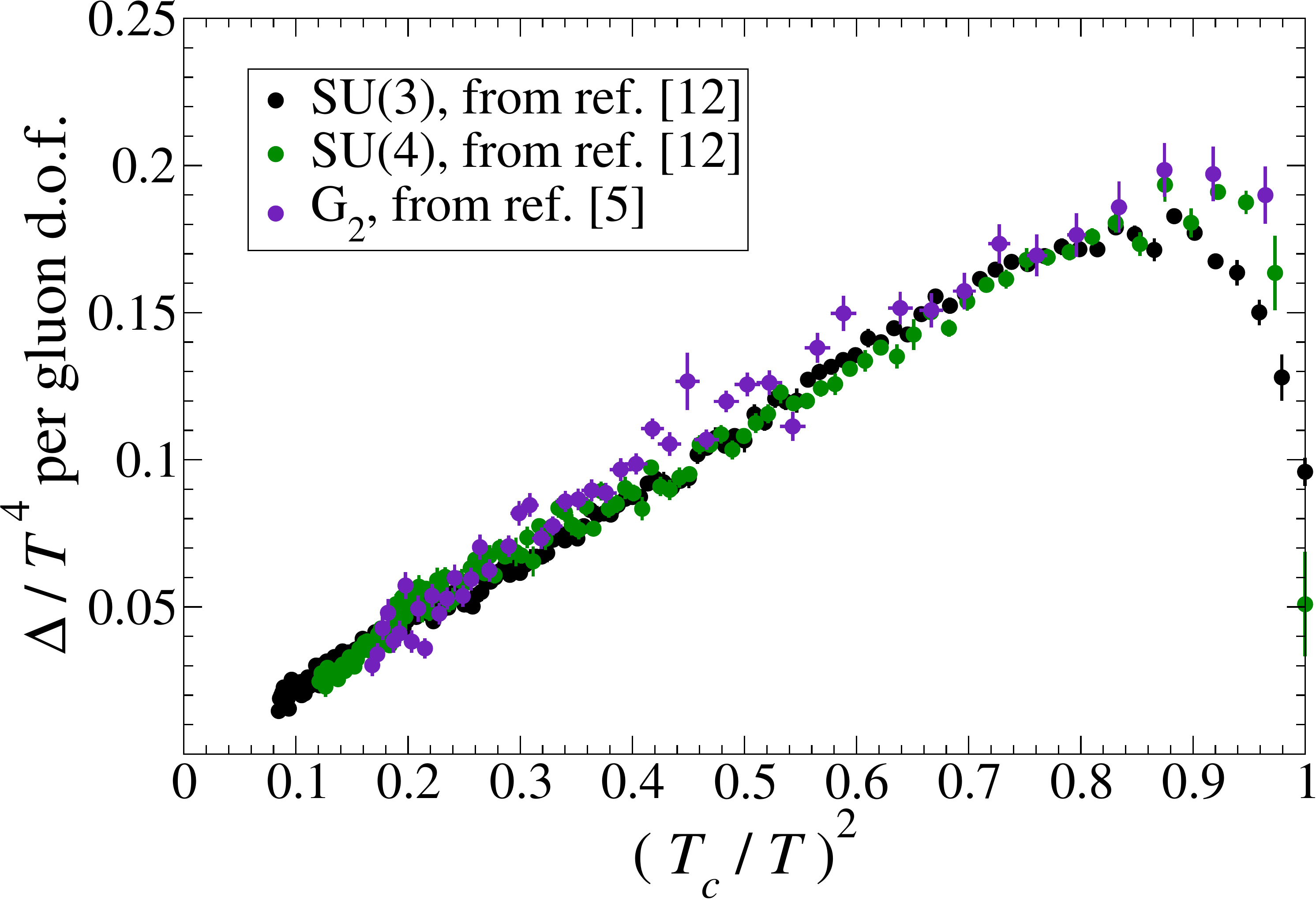}
\caption{\label{fig:g2_trace}Left panel: results for the trace anomaly for the $\Gtwo$ Yang-Mills theory obtained on lattices with different temporal extents.
Right panel: results for $\Delta$ per gluon degree of freedom in the deconfined phase for $\Gtwo$ (obtained on $N_t=6$ lattices) and for SU($N=3,4$) (taken from ref.~\cite{Panero:2009tv}), plotted on $(T_c/T)^2$.}
\end{figure}

In this section we review recent progress~\cite{Bruno:2014rxa} on the thermodynamics of the exceptional, center-less $\Gtwo$ gauge theory in the deconfined phase.
Results of lattice simulations for the trace anomaly $\Delta$ for different values of $N_t$ are showed in the left panel of figure~\ref{fig:g2_trace}.
The right panel of figure~\ref{fig:g2_trace} shows instead the value of $\Delta/T^4$ diveded by the number of gluon degrees of freedom (d.o.f.), i.e. by $2 \times d_a$: the factor 2 stands for the transverse polarizations of the gluon, while $d_a$ is the dimension of the adjoint representation, respectively $d_a=14$ for $\Gtwo$ and $d_a=N^2-1$ for SU($N$).

When expressed per gluon d.o.f., $\Gtwo$ results for $\Delta/T^4$ collapse on older SU($N$) (with $N=3,4$) data from ref.~\cite{Panero:2009tv}: indeed, the behaviour of these theories is very similar not only qualitatively but also quantitatively.
These findings strongly support the idea of universality for the thermal properties of different non-Abelian gauge theories~\cite{Lucini:2013qja}.
Furthermore, a rather peculiar behaviour, which was already observed for SU($N\geq3$) theories (see refs.~\cite{Borsanyi:2012ve,Panero:2009tv,Boyd:1996bx}), appears in the temperature range investigated in this work: the trace anomaly $\Delta$ presents nearly perfect proportionality to $1/T^2$: this suggests that non-perturbative effects are relevant in this temperature region, as the gluon plasma is not weakly coupled and cannot be described in purely perturbative terms.

\acknowledgments
I thank K.~K.~Szab{\'o} and the organizers of EPS-HEP~2015 for the invitation to present this contribution at the conference.

\end{document}